\documentclass[graybox,natbib]{svmult}

\usepackage{type1cm}        % activate if the above 3 fonts are
\usepackage{makeidx}         % allows index generation
\usepackage{graphicx}        % standard LaTeX graphics tool
\usepackage{multicol}        % used for the two-column index
\usepackage[bottom]{footmisc}% places footnotes at page bottom
\usepackage{todonotes}
\usepackage{booktabs}

\usepackage{framed,color}
\definecolor{shadecolor}{rgb}{.5,.5,.5} % change to suitable shade of grey

\usepackage{newtxtext}       % 
\usepackage{newtxmath}       % selects Times Roman as basic font

\makeindex             % used for the subject index
\begin{document}

\title*{The Evolution of Empirical Methods in Software Engineering}
% Use \titlerunning{Short Title} for an abbreviated version of
% your contribution title if the original one is too long
\author{Michael Felderer, Guilherme Horta Travassos}
% Use \authorrunning{Short Title} for an abbreviated version of
% your contribution title if the original one is too long
\authorrunning{Felderer and Travassos}
\institute{Michael Felderer \at University of Innsbruck, Austria, \\
Blekinge Institute of Technology, Sweden \\
\email{michael.felderer@uibk.ac.at} 
\and Guilherme Horta Travassos \at PESC/COPPE - Federal University of Rio de Janeiro, Brazil, \\ CNPq Researcher \\\email{ght@cos.ufrj.br}}
%
% Use the package ''url.sty'' to avoid
% problems with special characters
% used in your e-mail or web address
%
\maketitle

% deadline: 30th April

\abstract{Empirical methods like experimentation have become a powerful means to drive the field of software engineering by creating scientific evidence on software development, operation, and maintenance, but also by supporting practitioners in their decision-making and learning. Today empirical methods are fully applied in software engineering. However, they have developed in several iterations since the 1960s. In this chapter we tell the history of empirical software engineering and present the evolution of empirical methods in software engineering in five iterations, i.e., (1) mid-1960s to mid-1970s, (2) mid-1970s to mid-1980s, (3) mid-1980s to end of the 1990s, (4) the 2000s, and (5) the 2010s. We present the five iterations of the development of empirical software engineering mainly from a methodological perspective and additionally take key papers, venues, and books, which are covered in chronological order in a separate section on recommended further readings, into account. We complement our presentation of the evolution of empirical software engineering by presenting the current situation and an outlook in Section~\ref{sec:actual_situation_outlook} and the available books on empirical software engineering. Furthermore, based on the chapters covered in this book we discuss trends on contemporary empirical methods in software engineering related to the plurality of research methods, human factors, data collection and processing, aggregation and synthesis of evidence, and impact of software engineering research.}

\section{Introduction}
\label{sec:intro}

The term software engineering originated in the early 1960s~\citep{hey2014computing}. During the NATO Software Engineering Conferences held in 1968 and 1969, participants made explicit that engineering software requires dedicated approaches that are separate from those for the underlying hardware systems. Until that "software crisis", software-related research mostly focused on theoretical aspects, e.g., algorithms and data structures used to write software systems, or practical aspects, e.g., an efficient compilation of software for particular hardware systems~\citep{gueheneuc2019ese}. Since then, these topics are investigated in computer science, which pertains to understanding and proposing theories and methods related to the efficient computation of algorithms, and differs from software engineering (research), which has become a very dynamic discipline on its own since its foundation in the 1960s. \cite{ieee1990glossary,ieee2010vocabulary} defines software engineering (SE) as: (1) The application of a systematic, disciplined, quantifiable approach to the development, operation and maintenance of software; that is, the application of engineering to software, and (2) The study of approaches as in (1). Software engineering also differs from other engineering disciplines due to the immaterial nature of software not obeying physical laws and the importance of human factors as software is written by people for people. Software engineering is fundamentally an empirical discipline, where knowledge is gained applying direct and indirect observation or experience. Approaches to software development, operation, and maintenance must be investigated by empirical means to be understood, evaluated, and deployed in proper contexts. Empirical methods like experimentation are therefore essential in software engineering to gain scientific evidence on software development, operation, and maintenance, but also to support practitioners in their decision-making and learning~\citep{travassos2008environment}. The application of empirical methods makes software engineering more objective and less imprecise, facilitating the transfer of software technologies to the industry~\citep{shull2001empirical}. Software engineers learn by observing, exploring, and experimenting. The level of learning depends on the degree of observation or intervention~\citep{thomke2003experimentation} promoted by the experiences and studies performed.

Traditionally, \emph{empirical software engineering} (ESE) is the area of research that emphasizes the use of empirical methods in the field of software engineering. According to \cite{HarrisonBasiliESE}, "Empirical software engineering is the study of software-related artifacts for the characterization, understanding, evaluation, prediction, control, management, or improvement through qualitative or quantitative analysis. The quantitative studies may range from controlled experimentation to case studies. Qualitative studies should be well-defined and rigorous." The role and importance of the different types of empirical methods in software engineering have evolved since the foundation of software engineering. In this chapter, we discuss the evolution of empirical methods in software engineering and especially also take key venues and books into account as they reflect that evolution. 

This chapter is organized as follows: In Section~\ref{sec:emp_research_se}, we provide background on empirical research methods in software engineering. In Section~\ref{sec:evolution_ese}, we present the evolution of empirical software engineering by describing five iterations of its development. Based on that "historical" perspective on empirical software engineering, in Section~\ref{sec:actual_situation_outlook} we describe current trends in empirical software engineering based on the chapters on contemporary empirical methods in software engineering covered in this book. In Section~\ref{sec:recommended_reading}, we present the available books on empirical methods in software engineering in chronological order as recommended further reading. Finally, in Section~\ref{sec:conclusion}, we conclude this chapter.

\section{Empirical Research Methods in Software Engineering}
\label{sec:emp_research_se}

The scientific approach typically consists of observation, measurement, and experimentation. Observation helps researchers to formulate essential questions about a phenomenon under study to build models and to derive hypotheses that can be tested through experimentation. Measurement is essential for both observation and experimentation. A scientific hypothesis must be refutable to be meaningfully tested. Tested hypotheses are compiled and communicated in the form of laws or theories. At the heart of the scientific approach are research methods in general and the empirical method in particular. Empirical methods leverage evidence obtained through observation, measurement, or experimentation to address a scientific problem. Evidence should be based on qualitative and quantitative research. In this section, we provide an overview of research methods in software engineering in general and empirical methods in particular. 

\subsection{Research Methods}

To perform scientific research in software engineering, one has to understand the available research methods and their limitations. For the field of software engineering, \cite{basili1993experimental} and \cite{glass1994software} summarized four research methods: scientific, engineering, empirical, and analytical.

The so-called \emph{scientific method} observes the world and builds a model based on the observations, e.g., a simulation model of the software process or product. The scientific method is inductive and tries to extract from the world some model that can explain a phenomenon and to evaluate whether the model is representative for the phenomenon under observation. It is a model-building approach.

The \emph{engineering method} studies current solutions, proposes changes, and then evaluates them. It suggests the most appropriate solutions, develops, measures and analyzes, and repeats until no further improvement is possible. It is an evolutionary improvement-oriented approach that assumes the existence of some model of the software process or product. It modifies this model to improve the objects of study. 

The \emph{empirical method} proposes a model, and evaluates it through empirical studies like case studies or experiments. The empirical method normally follows an iterative and incremental approach that can begin with an exploratory survey, followed by case studies in an industrial context to better understand specific phenomena and controlled experiments to investigate cause-effect relationships. 

The \emph{analytical method} proposes a formal theory, develops the theory, derives the results, and, if possible, compares it with empirical observations. It is deductive and provides an analytical basis for developing a model.

Traditionally, the analytical method is used in the more formal areas of electrical engineering and computer science, but is important for software engineering as well, e.g., when building mathematical models for software reliability growth~\citep{lyu1996handbook}. The scientific method, inspired by natural science, is traditionally used in applied areas, such as the simulation of a sensors network to evaluate its performance. However, simulations are used as a means for conducting an experiment as well~\citep{BarrosWernerTravassos2004JSS}. The engineering method is dominating in industry~\citep{wohlin2012experimentation}. The empirical method, mainly using empirical strategies, has traditionally been used in social sciences and psychology, where one is unable to state any laws of nature but concerned with human behavior. The engineering and the empirical method can be seen as variations of the scientific method~\citep{basili1993experimental}. This overlap and an integrated view of the scientific, engineering, and empirical methods is also an underlying design principle of this book on empirical methods. It considers not only chapters on traditional empirical strategies like surveys (see Chapter "Challenges in Survey Research" in this book), but for instance, also a chapter on simulation-based studies (see Chapter "The Role of Simulation-Based Studies in Software Engineering Research"), which are closer to the scientific method as defined above, or a chapter on design science (see Chapter "The Design Science Paradigm as a Frame for Empirical Software Engineering"), which can tightly be linked to the engineering method. All of these investigation strategies refer to empirical methods.

\subsection{Empirical Methods}

Empirical methods rely on the collected data. Data collection methods may involve qualitative or quantitative data. Some widely used \emph{qualitative data collection methods} in software engineering are interviews and participant observation~\citep{seaman1999qualitative}. Some commonly used \emph{quantitative data collection methods} are archival data, surveys, experiments, and simulation~\citep{wohlin2012experimentation}. Once data are collected, the researcher needs to analyze the data by using \emph{qualitative analysis methods}, e.g., grounded theory, thematic analysis or hermeneutics, and \emph{quantitative analysis methods}, e.g., statistical analysis and mathematical modeling approaches.

In general, there are three widely recognized research processes called quantitative research, qualitative research, and semiquantitative research. An alternative option is the combination of both qualitative and quantitative research, denoted as mixed research~\citep{creswell2018research}. The distinction between qualitative and quantitative research comes not only from the type of data collected, but also the objectives, types of research questions posed, analysis methods, and the degree of flexibility built into the research design as well~\citep{wohlin2015towards}. \emph{Qualitative research} aims to understand the reason (i.e., "why") and mechanisms (i.e., "how") explaining a phenomenon. A popular method of qualitative research is case study research, which examines a set of selected samples in detail to understand the phenomenon illustrated by the samples. For instance, a qualitative study can be conducted to understand the impediments of automating system tests. \emph{Quantitative research} is a data-driven approach used to gain insights about an observable phenomenon. Data collected from observations are analyzed using mathematical and statistical models to derive quantitative relationships between different variables capturing different aspects of the phenomenon under study. A popular method of quantitative research are controlled experiments to examine cause-effect relationships between different variables characterizing a phenomenon in a controlled environment. For instance, different review techniques could be compared via a controlled experiment. \emph{Mixed research} collects quantitative and qualitative data. It is a particular form of \emph{multi-method research}, which combines different research methods to answer some hypotheses, and is often used in empirical software engineering due to the lack of theories in software engineering with which we interpret quantitative data and due to the need to discuss qualitatively the impact of the human factor on any experiments in software engineering~\citep{gueheneuc2019ese}. \emph{Semiquantitative research} deals with approximate measurements to data rather than exact measurements~\citep{bertinsemiquantitavive}. It looks for understanding the behavior of a system based on causal relations between the variables describing the system. Semiquantitative models allow one to express what is known without making inappropriate assumptions, simulating ranges of behavior rather than values of point~\citep{Widman1989Semi}. It has many applications in both the natural and social sciences. Semiquantitative research supports cases where direct measurements are not possible, but where it is possible to estimate an approximated behavior. In other words, this type of study is applied in scenarios where the numerical values in the mathematical relations governing the changes of a system are not known. In this context, the direction of change is known, but not the size of its effect~\citep{OgbornMiller1994Semi}. Simulation-based studies in software engineering can benefit from using semiquantitative research~\citep{maraujosimulation}.

The three major and well-established empirical methods in software engineering are: survey, case study, and experiment~\citep{wohlin2012experimentation}. Primary studies using such methods can be performed in vivo, in vitro, in virtuo, and in silico~\citep{travassos2003contributions}. \emph{In vivo} studies involve participants and projects in their natural environments and contexts. Usually executed in software development organizations throughout the software development process under real working conditions. \emph{In vitro} studies are performed in controlled environments, such as laboratories or controlled communities, under configured working conditions. \emph{In virtuo} studies have the subjects interacting with a computerized model of reality. The behavior of the environment with which subjects interact is described as a model and represented by a computer program. \emph{In silico} studies represent both subjects and real world as computer models. The environment is fully composed of computer models to which human interaction is reduced to a minimum.

A \emph{survey} is a system for collecting information from or about subjects (people, projects, among others) to describe, compare, or explain their knowledge, attitudes, and behavior~\citep{fink2003survey}. A survey is often an investigation performed in retrospect, when, for instance, a tool or technique, has been in use for a while~\citep{pfleeger1995experimental}. The primary means of gathering qualitative or quantitative data are interviews or questionnaires. These are done through taking a sample that is representative of the population to which is generalized. 

A \emph{case study} in software engineering is an empirical inquiry that draws on multiple sources of evidence to investigate one or a small number of instances of a contemporary software engineering phenomenon within its real-life context, especially when the boundary between phenomenon and context cannot be clearly specified~\citep{runeson2012case}.

An \emph{experiment} is used to examine cause-effect relationships between different variables characterizing a phenomenon~\citep{gueheneuc2019ese}. Experiments allow researchers to verify, refute, or validate hypotheses formulated about the phenomenon under study. In a controlled experiment, one variable of the study setting is manipulated, and based on randomization, different treatments are applied to or by different subjects while keeping other variables constant, and measuring the effects on outcome variables~\citep{wohlin2012experimentation}. A quasi-experiment is similar to a controlled experiment, where the assignment of treatments to subjects cannot be based on randomization, but emerges from the characteristics of the subjects or objects themselves~\citep{wohlin2012experimentation}. Replication experiments reproduce or quasi-reproduce previous experiments with the objectives to confirm or infirm the results from previous experiments or to contrast previous results in different contexts~\citep{gueheneuc2019ese}. 

Regardless of the applied empirical method, to acquire scientific evidence about the investigated software engineering phenomena involves observation, measurement, and experimentation of the world and existing solutions. It demands the proposition of models and theories describing the observed behavior, collecting and analyzing data, putting the hypotheses under proof, and repeating the overall process over time to strengthen the evidence on the observed phenomena. Based on several primary studies, in which direct observations and measurements about the objects of interest are made, whether by surveys, experiments, or case studies, which are there also called \emph{empirical strategies}, one can perform secondary studies. A \emph{secondary study} does not generate any data from direct observation or measurement, instead, it analyzes a set of primary studies and usually seeks to aggregate the results from these to provide stronger forms of evidence about a particular phenomenon~\citep{kitchenham2015evidence}. Secondary studies typically appear as \emph{systematic (literature) reviews}, which aim to provide an objective and unbiased approach to finding relevant primary studies, and for extracting, aggregating and synthesizing the data from these~\citep{kitchenham2015evidence}. A particular type of a systematic review is a \emph{systematic mapping study}~\citep{petersen2015guidelines}, which classifies studies to identify clusters of studies (that could form the basis of a fuller review with more synthesis) and gaps indicating the need for more primary studies.

The scientific or industrial significance of empirical studies depends on their validity, i.e., the degree to which one can trust the outcomes of an empirical study~\citep{kitchenham2015evidence}. Validity is usually assessed in terms of four commonly encountered forms of \emph{threats to validity}: internal, external, construct, and conclusion validity~\citep{shadish2002experimental}. \emph{Internal validity} refers to inferences that the observed relationship between treatment and outcome reflects a cause-effect relationship. \emph{External validity} refers to whether a cause-effect relationship holds over other conditions, including persons, settings, treatment variables, and measurement variables. \emph{Construct validity} refers to how concepts are operationalized as experimental measures. \emph{Conclusion validity} refers to inferences about the relationship between treatment and outcome variables. 

The accomplishment of empirical studies relies on performing well-defined and evolutionary activities. The classical empirical study process consists of five phases: definition, planning, operation, analysis, and interpretation, as well as reporting and packaging~\citep{juristo2001basics,malhotra2016empirical}. The definition phase makes the investigated problem and overall objectives of the study explicit. The planning phase covers the study design and includes the definition of research questions and hypotheses as well as the definition of data collection, data analysis, and validity procedures. In the operation phase, the study is actually conducted. In the analysis and interpretation phase, the collected data is analyzed, assessed, and discussed. Finally, in the reporting and packaging phase, the results of the study are reported (e.g., in a journal article, a conference paper, or a technical report) and suitably packaged to provide study material and data. The latter has become more critical recently due to the open science movement (see Chapter "Open Science in Software Engineering" in this book).

\section{Evolution of Empirical Software Engineering}
\label{sec:evolution_ese}

The application of empirical methods in general and empirical software engineering in particular is well established in software engineering research. Almost all papers published in major software engineering venues these days include an empirical study~\citep{theisen2017writing}. Furthermore, since 2000, research methodology has received considerable attention in the software engineering research community resulting in many available publications on empirical research methodology in software engineering. In a recent mapping study, \cite{molleri2019cerse} identified 341 methodological papers on empirical research in software engineering.

The application of empirical methods and the underlying research methodology has developed iteratively since the foundation of software engineering in the 1960s. \cite{gueheneuc2019ese} discuss landmark articles, books, and venues in empirical software engineering that indicate the iterative development of the field. \cite{bird2015art} distinguish four "generations" of analyzing software data, i.e., preliminary work, academic experiments, industrial experiments, and "data science everywhere." In this section, we present five iterations of the development of empirical software engineering from a methodological perspective. We additionally take articles and venues into account, which is needed for a holistic understanding of the field's development. We complement our presentation of the evolution of empirical software engineering by presenting the current situation and an outlook in Section~\ref{sec:actual_situation_outlook} and the available books on empirical software engineering in chronological order in Section~\ref{sec:recommended_reading} on recommended further reading.

\subsection{First Iteration: Mid-1960s to Mid-1970s}

In the early years of software engineering, empirical studies were rare, and the only research model commonly in use was the analytical method, where different formal theories were advocated devoid of any empirical evaluation~\citep{glass1994software}. According to a systematic literature review of empirical studies performed by \cite{zendler2001preliminary}, \cite{grant1967exploratory} published the first empirical study in software engineering in 1967. The authors conducted an experiment that compared the performance of two groups of developers, one working with online access to a computer through a terminal and the other with offline access in batch mode. Another empirical study published early in the history of software engineering was an article by \cite{knuth1971empirical}, in which the author studied a set of Fortran programs to understand what developers do in Fortran programs. \cite{akiyama1971example} describes the first known "size law"~\citep{bird2015art}, stating that the number of defects is a function of the number of lines of code. The authors in these and other early studies defined the goal of the study, the questions to research, and the measures to answer these questions in an ad hoc fashion~\citep{gueheneuc2019ese}. However, they were pioneers in the application of empirical methods in software engineering.

\subsection{Second Iteration: Mid-1970s to Mid-1980s}

In the second iteration, already more empirical studies, mainly in-vitro experiments, were conducted. Prominent examples are experiments on structured programming~\citep{lucas1976structured}, flowcharting~\citep{shneiderman1977experimental}, and software testing~\citep{myers1978controlled}. The second iteration is characterized by first attempts to provide a systematic methodology to define empirical studies in software engineering in general and experiments in particular. These attempts culminated in the definition of the Goal/Question/Metrics (GQM) approach by~\cite{basiliweiss1984GQM}. The GQM approach helped practitioners and researchers to define measurement programs based on goals related to products, processes, and resources that can be achieved by answering questions that characterize the objects of measurement using metrics. The methodology has been used to define experiments in software engineering systematically.

In that iteration, empirical software engineering was also institutionalized for the first time. In 1976, the NASA Goddard Software Engineering Laboratory (NASA/SEL) was established at the University of Maryland, College Park (USA), aiming at support the observation and understanding of software projects~\citep{BasiliZelkowitzz2007CS}. The establishment of NASA/SEL provided the means to strengthen the importance of using basic scientific and engineering concepts in the context of software engineering~\citep{McGarrySoftwareProcess1994}. The paradigm change provided by using GQM~\citep{basiliweiss1984GQM}, including the ability of packaging knowledge on how to better build a software system, improved the way experiences could be organized and shared. The building and evolution of models at NASA/SEL pave the road for organizing the Experience Factory model~\citep{BasiliCaldieraRombachEF} and the dissemination of initial good practices on empirical software engineering.

\subsection{Third Iteration: Mid-1980s to End of the 1990s}

In the third iteration, not only experiments but also surveys (for instance, by~\cite{burkhard1989applications} on the application of computer-aided software engineering tools) and case studies (for instance, by~\cite{curtis1988field} on the software design process for large systems) were performed to some extent. Also, the explicit discussion of threats to validity appeared in that iteration. One of the first studies explicitly discussing its threats to validity was an article by \cite{swanson1988use} on the use of case study data in software management research. From the late 1980s, researchers also started to analyze software data using algorithms taken from artificial intelligence research~\citep{bird2015art}. For instance, decision trees and neural networks were applied to predict error-proneness~\citep{porter1990empirically}, to estimate software effort~\citep{srinivasan1995machine} and to model reliability growth~\citep{tian1995integrating}.

In the third iteration, empirical studies began to attract the attention of several research groups all over the world, who realized the importance of providing empirical evidence about the developed and investigated software products and processes. The experiences shared by NASA/SEL and the participation of several researchers in conducting experiments together with NASA/SEL helped to strengthen the use of different experimental strategies and the application of surveys.

The interest in the application of the scientific method by different researchers, the identification of the need to evolve the experimentation process through sharing of experimental knowledge among peers as well as the transfer of knowledge to industry, among other reasons, led to the establishment of the International Software Engineering Research Network (ISERN) in 1992. ISERN held its first annual meeting in Japan in 1993 sponsored by the Graduate School of Information Science at the Nara Institute of Science and Technology.

The need to share the ever increasing number of studies and their results, and the growing number of researchers applying empirical methods in software engineering lead to the foundation of suitable forums. In 1993 the IEEE International Software Metrics Symposium, in 1996, the Empirical Software Engineering International Journal, and 1997, the Empirical Assessments in Software Engineering (EASE) event at Keele University were founded.

By the end of this iteration, several institutes dedicated to empirical software engineering were established. In 1996, the Fraunhofer Institute for Experimental Software Engineering (IESE) associated with the University of Kaiserslautern (Germany) was established. In 1998, the Fraunhofer Center for Experimental Software Engineering (CESE) associated with the University of Maryland, College Park (USA) began operations. Also, other institutions and laboratories, such as National ICT Australia as well as the Simula Research Laboratory and SINTEF (both located in Norway), among others, started to promote empirical studies in software engineering in the industry.

Finally, by the end of the 1990s, the publication of methodological papers on empirical methods in software engineering started. \cite{zelkowitz1998experimental} provided an overview of experimental techniques to validate new technologies, \cite{seaman1999qualitative} provided guidelines for qualitative data collection and analysis, and \cite{basili1999building} discussed families of experiments.

\subsection{Fourth Iteration: The 2000s}

Since 2000 research methodology has received considerable attention, and therefore the publication of methodological papers further increased. For instance, \cite{host2000using} discuss the usage of students as subjects in experiments, \cite{shull2001empirical} describe a methodology to introduce software processes based on experimentation, \cite{pfleeger2001principles} provide guidelines on surveys in software engineering, \cite{lethbridge2005studying} provide a classification of data collection methods, \cite{KitchenhamCharters2007SLR} provide guidelines for performing systematic literature reviews in software engineering, \cite{shull2008role} discuss the role of replication in empirical software engineering, and \cite{runeson2009guidelines} provide guidelines for case study research. In connection to the increased interest in research methodology, also the first books on empirical research methods in software engineering with a focus on experimentation written by \cite{wohlin2000introduction} and \cite{juristo2001basics} appeared around 2000 (see Section~\ref{sec:recommended_reading} for a comprehensive overview of books on empirical software engineering). Also, combining research methods and performing multi-method research became more popular in the period. One of the first papers following a multi-method research methodology was published by \cite{espinosa2002shared} on shared mental models, familiarity, and coordination in distributed software teams.

With the growing number of empirical studies, knowledge aggregation based on these primary studies became more crucial to understand software engineering phenomena better. No single empirical study on a software engineering phenomenon can be considered definitive~\citep{shull2004knowledge} and generalized to any context. Therefore, the replication of studies in different contexts is of paramount importance to strengthen its findings. However, the existence of conclusive, no conclusive, contradictory, and confirmatory results about a particular software engineering phenomenon should be combined to strengthen the evidence on software phenomena or to reveal the need for more primary studies on the phenomenon of interest. In consequence, there arose a need for secondary studies that aim to organize, aggregate, and synthesize all relevant results from primary studies regarding a particular phenomenon under research. \cite{Kitchenham2004SLR} was the first who recommended the use of systematic literature reviews (SLRs) in software engineering and adapted respective guidelines, mainly from medical research, to software engineering. With the guidelines of~\cite{Kitchenham2004SLR} and \cite{Biolchini2005SLR}, the empirical software engineering community had a tool to systematically synthesize knowledge available in primary studies, which spread rapidly and enabled evidence-based software engineering~\citep{kitchenham2004evidence}. In a systematic review of SLRs in software engineering, \cite{kitchenham2013systematic} identified 68 papers reporting 63 unique SLRs published in SE conferences and journals between 2005 and mid-2012. \cite{petersen2008systematic} clarify and expand upon the differences between SLRs and systematic mapping studies, and provide guidelines for the latter. In their seminal paper on the future of empirical methods in software engineering research, \cite{sjoberg2007future} present the important role of synthesis of empirical evidence in their vision of software engineering research. The vision is that for all fields of software engineering, empirical research methods should enable the development of scientific knowledge about how useful different SE technologies are for different kinds of actors, performing different kinds of activities, on different kinds of systems to guide the development of new SE technology and important SE decisions in industry. Major challenges to the pursuit of this vision are more and better synthesis of empirical evidence, and connected to that building and testing more theories as well as increasing quality, including relevance, of studies.

One of the problems faced by the software engineering community has often been the scarcity of software data for conducting empirical studies~\citep{malhotra2016empirical}. The availability of (open) source code repositories and software process data due to automated or even continuous software engineering enabled new data mining approaches in software engineering in that period. In a seminal paper, \cite{zimmermann2005mining} used association rule learning to find patterns of defects in a large set of open-source projects. Furthermore, also, software data from companies were analyzed. For instance, at AT\&T, \cite{ostrand2004bugs} code metrics to predict defects, and at Microsoft -- which even founded an own Empirical Software Engineering (ESE) group in Microsoft Research~\citep{bird2011empirical} -- \cite{nagappan2005use} showed that data from that organization could predict software quality. In consequence, also repositories -- like the PROMISE repository -- that collect software data and make them publicly available were founded. The PROMISE repository was founded in 2005 and seeded with NASA data~\citep{menzies2014sharing}. The empirical evidence gathered through analyzing the data collected from the software repositories is considered to be an important support for the (empirical) software engineering community these days. There are even venues that focus on analysis of software data such as Mining Software Repositories (MSR), which was organized for the first time in 2004 in Edinburgh (UK) and Predictive Models and Data Analytics in Software Engineering (PROMISE), which was organized for the first time in 2005 in St. Louis (USA).

In general, the growing interest in empirical software engineering in that period resulted in projects such as the Experimental Software Engineering Research Network (ESERNET) in Europe from 2001 to 2003 and the foundation of several venues. In 2007, the first ACM/IEEE International Symposium on Empirical Software Engineering and Measurement (ESEM) was held in Madrid (Spain). ESEM is the result of the merger between the ACM/IEEE International Symposium on Empirical Software Engineering, which ran from 2002 to 2006, and the IEEE International Software Metrics Symposium, which ran from 1993 to 2005. In 2003, Experimental Software Engineering Latin American Workshop (ESELAW) was organized for the first time. Also, in 2003, the International Advanced School on Empirical Software Engineering (IASESE) performed its first set of classes in Rome (Italy). In 2006, the International Doctoral Symposium on Empirical Software Engineering (IDoESE) was founded. Today, the ISERN annual meeting, IASESE, IDoESE, and ESEM form the Empirical Software Engineering International Week (ESEIW), which is held annually.

\subsection{Fifth iteration: The 2010s}

Since 2010 empirical studies are "everywhere" in software engineering. Almost all papers in major software engineering conferences like ICSE contain empirical studies. Also, more and more books dedicated to empirical research methodology in software engineering are published (see Section~\ref{sec:recommended_reading}), and papers on empirical research methodology are published at a constant pace. For instance, \cite{ivarsson2011method} present a model for evaluating the rigor and relevance of technology evaluations in industry, \cite{arcuri2014hitchhiker} provide a guide to statistical tests for assessing randomized algorithms in software engineering, \cite{wieringa2014empirical} discusses scaling up of empirical methods for technology validation in practice, \cite{wohlin2015towards} provide a decision-making structure for selecting a research design, \cite{mellosampling} provide probabilistic sampling approaches for large-scale surveys, \cite{sharp2016role} discuss the use and value of ethnographic studies in software engineering research, \cite{stol2016grounded} discuss the use of grounded theory and their reporting, \cite{briand2017case} discuss the importance of context and the overrating of generalizability in software engineering, and \cite{stol2018abc} provide a holistic framework for software engineering research. Furthermore, \cite{harman2010search} provide a comprehensive overview and guidance on the application of search-based optimization in software engineering. Especially, in this period many papers presenting results on search-based software engineering, that generally (though not exclusively) fall in the category of empirical software engineering papers were published. Due to the potentially high computational complexity of optimization algorithms, some researchers have started to use high performance computing environments to support the execution of their studies~\citep{farzathpc}.

In this iteration, one can observe a growing interest in the role of theory within software engineering research to develop the field further as a scientific discipline. In December 2009, the Software Engineering Method and Theory (SEMAT) initiative was launched that aims towards the development of a general theory of software engineering. SEMAT organized several events, among others, a workshop series on a General Theory of Software Engineering (GTSE) between 2012 and 2015. \cite{stol2015theory} even argue for a theory-oriented software engineering research perspective, which can complement the recent focus on evidence-based software engineering. Also, several concrete theories have been developed in that iteration. For instance, \cite{johnson2016tarpit} present a general theory of software engineering called Tarpit, \cite{bjarnason2016theory} a theory of distances in software engineering, and \cite{wagner2019status} a theory on requirements engineering.

Today not only almost all papers in major software engineering conferences contain empirical studies, but also most software engineering conferences have explicitly integrated empirical software engineering into their program, e.g., as dedicated sessions or tracks. In addition, there are several workshops on conducting empirical studies in specific areas. For instance, at ICSE, there has been a collocated International Workshop on Conducting Empirical Studies in Industry (CESI) and at RE the International Workshop on Empirical Requirements Engineering (EmpiRE). The Experimental Software Engineering Latin American Workshop (ESELAW) joined the Ibero-American Conference on Software Engineering (CIbSE) in 2011 as a colocated workshop and became a dedicated track in 2013 due to the increased number of submissions.

\begin{figure}
    \centering
    \includegraphics[width=\textwidth]{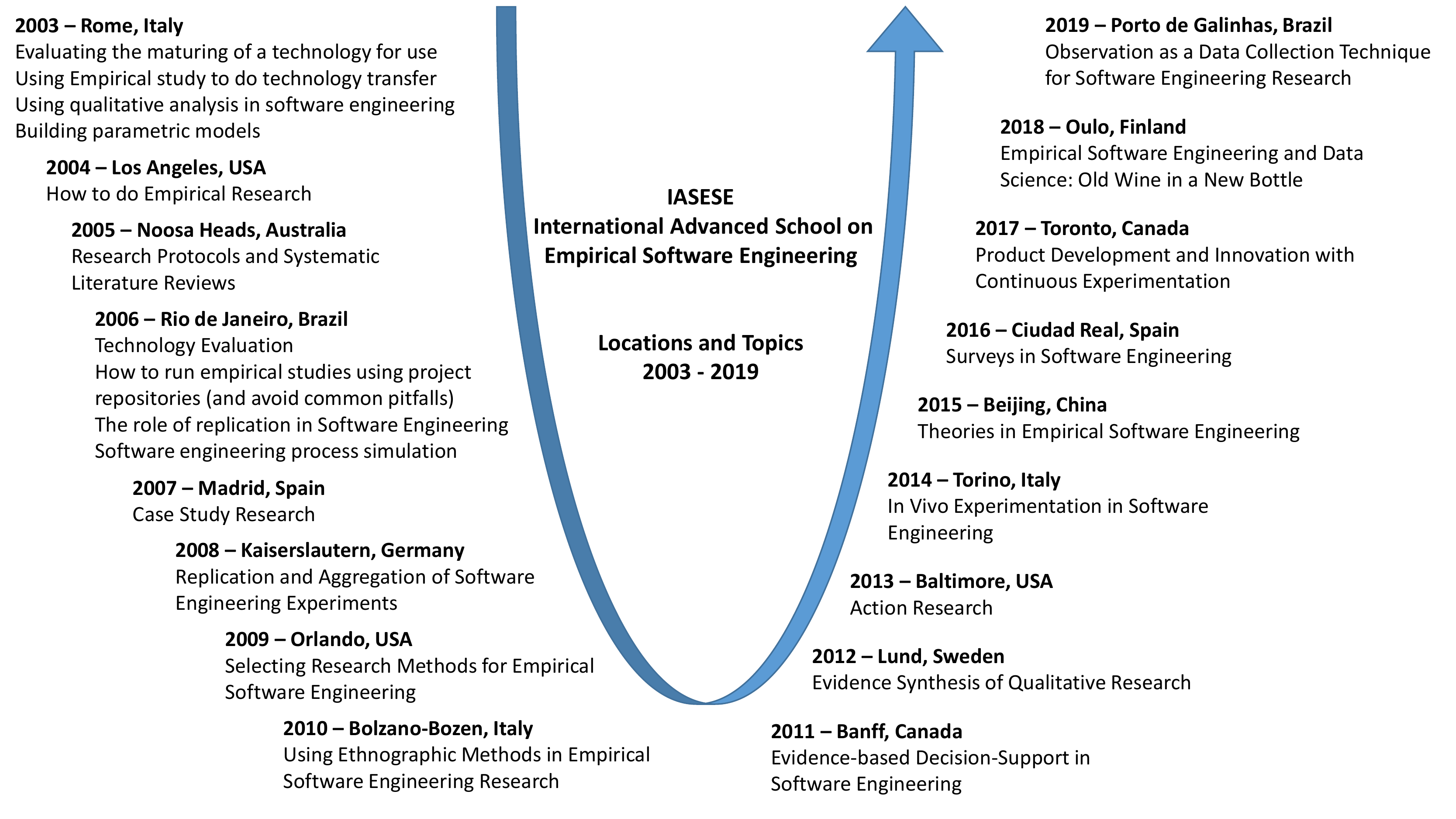}
    \caption{International Advanced School on Empirical Software Engineering (IASESE) Timeline and Topics from 2003 to 2019}
    \label{fig:iasese-timeline}
\end{figure}

ESEIW (including ESEM) and EASE are established as the leading annual events to discuss methodological issues on empirical research in software engineering. Empirical methods have been an explicit topic in several summer schools including the annual LASER summer school (which hosted the topic empirical software engineering in 2010), PASED - Canadian Summer School on Practical Analyses of Software Engineering Data in 2011, the Empirical Research Methods in Software Engineering and Informatics (ERMSEI) in 2016 and 2017, the International Summer School on Software Engineering (SIESTA) in 2018 and 2019 as well as the 2019 Summer School in Empirical Software Engineering at Brunel (UK). In the context of ESEIW, the International Advanced School on Empirical Software Engineering (IASESE) has been organized annually since 2003 and helped to spread knowledge on current empirical methods in software engineering among junior and senior researchers. Figure~\ref{fig:iasese-timeline} presents the IASESE timeline and its topics along the places and years. The topics taught over the years also reflect the evolution of empirical software engineering, as discussed in this section.

\section{Current Situation and Outlook}
\label{sec:actual_situation_outlook}

Since the first empirical studies in the 1960s, the field of empirical software engineering has considerably matured in several iterations. However, the empirical methods resulting from the five iterations presented in the previous section are not the end of the story, and as in any scientific discipline, research methods develop further. The chapters of this book discuss contemporary empirical methods that impact the current evolution of empirical software engineering and form the backbone of its next iteration. For sure, the description of the current situation and future trends is never complete and always subjective to some extent. But we think that the chapters covered in this book show several interesting trends in contemporary empirical methods in software engineering, which we want to summarize here. 

The evolution of empirical software engineering leads to the continuous adoption of empirical methods from other fields and the refinement of existing empirical methods in software engineering. The resulting plurality of research methods requires guidance in knowledge-seeking and solution-seeking (i.e., design science) research. The Chapter "Guidelines for Conducting Software Engineering Research" presents guidelines for conducting software engineering research based on the \emph{ABC framework}, where ABC represents the three desirable aspects of research generalizability over actors (A), precise control of behavior (B), and realism of context (C). Each empirical method has its strengths and weaknesses. It is beneficial to utilize a mix of methods depending on the research goal or even to combine methods. Case survey research combines case study and survey research, which rely primarily on qualitative and quantitative data, respectively. The Chapter "Guidelines for Case Survey Research in Software Engineering" provides an overview of the \emph{case survey method}. While being an important and often used empirical method, survey research has been less discussed on a methodological level than other types of empirical methods. The Chapter "Challenges in Survey Research" discusses \emph{methodological issues in survey research} for software engineering concerning theory building, sampling, invitation and follow-up, statistical analysis, qualitative analysis, and assessment of psychological constructs. Although software engineering is an engineering discipline, the design science paradigm has been explicitly adapted to software engineering relatively late by~\cite{wieringa2014design}, and the full potential of the design science paradigm has not been exploited so far in software engineering. The Chapter "The Design Science Paradigm as a Frame for Empirical Software Engineering" uses the \emph{design science paradigm} as a frame for empirical software engineering and uses it to support the assessment of research contributions, industry-academia communication, and theoretical knowledge building.

It is generally acknowledged that software development is a human-intensive activity as software is built by humans for humans. However, traditionally SE research has focused on artifacts and processes without explicitly taking human factors in general and the developer perspective in particular into account. If the perspective on how developers work was considered, then it was mostly measured from a subjective perspective, e.g., by interviews or opinion surveys, or a "black-box" perspective by mining repository data or measuring the created development artifacts. The Chapter "Biometric Measurement in Software Engineering" introduces \emph{biometric sensors and measure} that provide new opportunities to more objectively measure physiological changes in the human body that can be linked to various psychological processes. These biometric measurements can be used to gain insights on fundamental cognitive and emotional processes of software developers while they are working, but also to provide better and more prompt tool support for developers. Another human-related issue is the involvement of humans in empirical studies, especially in experiments. On the one hand, it normally difficult to recruit a significant number of professionals for an empirical study, and on the other hand, measurements are invasive. The Chapter "Empirical Software Engineering Experimentation with Human Computation" explores the potential of \emph{human computation methods}, such as crowdsourcing, for experimentation in empirical software engineering.

Empirical methods rely on the collected data. However, the volume, velocity, and variety of data in software products and processes have exploded during the last years. Therefore, the new scientific paradigm of \emph{data science} has gained much attention, also within software engineering. The Chapter "Data Science and Empirical Software Engineering" relates to traditional ESE and data science. It shows that both paradigms have many characteristics in common and can benefit from each other. Given large data sets, \emph{optimization} is an important form of data analytics support of human decision-making. Empirical studies serve both as a model and as data input for optimization. The Chapter "Optimization in Software Engineering: A Pragmatic Approach" provides an overview of optimization in software engineering in general and its value and applicability in ESE in particular. With increased automation, uncertainty (due to the application of statistical models), and monitoring capabilities in data-driven software engineering, also the role of \emph{simulation} techniques becomes more important. The Chapter "The Role of Simulation-Based Studies in Software Engineering Research" provides a guide to simulation-based studies in software engineering. \emph{Bayesian data analysis} is a means to embrace uncertainty by using multilevel statistical models and making use of all available information at hand. The Chapter "Bayesian Data Analysis in Empirical Software Engineering: The Case of Missing Data" provides an introduction to Bayesian data analysis and an application example to empirical software engineering dealing with common issues in ESE like missing data. 

Extracting, aggregating, and synthesizing evidence from empirical studies is essential for the development of scientific knowledge and the field of software engineering. However, conducting secondary studies like systematic literature reviews and aggregating evidence is still challenging. Conducting systematic literature reviews (SLRs) is largely a manual and, therefore, time-consuming and error-prone process. The Chapter "Automating Systematic Literature Review" provides strategies to \emph{automate the SLR process}. Secondary studies often lack connection to software engineering practice, which is essential to software engineering. The Chapter "Rapid Reviews in Software Engineering" presents the concept of \emph{rapid reviews}, which are lightweight secondary studies focused on delivering evidence to practitioners on time. Another approach to link to practice is to take grey literature into account in empirical studies. The Chapter "Benefitting from Grey Literature in Software Engineering Research" discusses the concept of \emph{grey literature} in software engineering and ways how to consider it in primary and secondary studies. Considering that secondary studies are often used to support the evidence-based paradigm, it is crucial to managing their threats properly. The Chapter "Guidelines for Managing Threats to Validity of Secondary Studies in Software Engineering" provides guidelines for managing \emph{threats to validity of secondary studies} in software engineering. Evidence in software engineering is often rare and produced in both quantitative and qualitative forms. It makes the synthesis of evidence, which is an essential element in scientific knowledge creation, a challenging task. The Chapter "Research Synthesis in Software Engineering" provides an overview of research \emph{synthesis methods} in software engineering.

Society in general and funding agencies, in particular, put a stronger focus on the impact of (software engineering) research. Therefore, open science and research transfer are becoming essential topics in (empirical) software engineering. \emph{Open science} describes the movement of making any research artifact available to the public and includes open access, open data, and open source. The topic is natural and especially important in empirical software engineering to guarantee the replicability of empirical studies. The Chapter "Open Science in Software Engineering" reflects upon the essentials in open science for software engineering to help to establish a common ground and to make open science a norm in SE. Industry-academia collaboration is one of the cornerstones of empirical software engineering. However, close and sustainable collaboration with industry are key issues in the field. The Chapter "Third Generation Industrial Co-production in Software Engineering" presents a seven-step \emph{industrial co-production} approach that enables deep and long-term industry-academia collaboration.

\section{Recommended Further Reading}
\label{sec:recommended_reading}

Since 2000 research methodology has received considerable attention in the software engineering research community. Therefore, plenty of literature is available on empirical research methodology in software engineering. \cite{molleri2019cerse} identified in a recent systematic mapping study 341 methodological papers on empirical research in software engineering - and therefore, a complete overview would exceed the scope of this book chapter. However, following the style of this book chapter, we provide an overview of the available English text and special issue books explicitly dedicated to empirical research methodology in software engineering in chronological order of their publication.

\cite{wohlin2000introduction} published a book entitled ''Experimentation in Software Engineering'', which provides an overview of the core empirical strategies in software engineering, i.e., surveys, experimentation and case studies and as its main content all steps in the experimentation process, i.e., scoping, planning, operation, analysis and interpretation as well as presentation and package. The book is complemented by exercises and examples, e.g., an experiment comparing different programming languages. Consequently, the book targets students, teachers, researchers, and practitioners in software engineering. In 2012 a revision of this popular book had been published with Springer~\citep{wohlin2012experimentation}.

\cite{juristo2001basics} published a book entitled ''Basics of Software Engineering Experimentation'', which presents the basics of designing and analyzing experiments both to software engineering researchers and practitioners based on SE examples like comparing the effectiveness of defect detection techniques. The book presents the underlying statistical methods, including the computation of test statistics in detail.

\cite{endres2003handbook} published ''A handbook of software and systems engineering. Empirical observations, laws, and theories''. The book presents rules, laws, and their underlying theories from all phases of the software development lifecycle. The book provides the reader with clear statements of software and system engineering laws and their applicability as well as related empirical evidence. The consideration of empirical evidence distinguishes the book from other available handbooks and textbooks on software engineering. 

\cite{juristo2003lecture} edited ''Lecture notes on empirical software engineering'', which aims to spread the idea of the importance of empirical knowledge in software development from a highly practical viewpoint. It defines the body of empirically validated knowledge in software development to advise practitioners on what methods or techniques have been empirically analyzed and what the results were. Furthermore, it promotes ''empirical tests'', which have traditionally been carried out by universities or research centers, for application in industry to validate software development technologies used in practice.

\cite{shull2007guide} published the ''Guide to advanced empirical software engineering''. It is an edited book written by experts in empirical software engineering. It covers advanced research methods and techniques, practical foundations, as well as knowledge creation, approaches. The book at hand provides a continuation of that seminal book covering recent developments in empirical software engineering. 

\cite{runeson2012case} published a book entitled ''Case study research in software engineering: Guidelines and examples'', which covers guidelines for all steps of case study research, i.e., design, data collection, data analysis and interpretation, as well as reporting and dissemination. The book is complemented with examples from extreme programming, project management, quality monitoring as well as requirements engineering and additionally also provides checklists.

\cite{wieringa2014design} published a book entitled ''Design science methodology for information systems and software engineering'', which provides guidelines for practicing design science in software engineering research. A design process usually iterates over two activities, i.e., first designing an artifact that improves something for stakeholders, and subsequently empirically validating the performance of that artifact in its context. This ''validation in context'' is a key feature of the book.

\cite{menzies2014sharing} published a book entitled ''Sharing data and models in software engineering''. The central theme of the book is how to share what has been learned by data science from software projects. The book is driven by the PROMISE (Predictive Models and Data Analytics in Software Engineering) community. It is the first book dedicated to data science in software and mining software repositories. Closely related to this book, \cite{bird2015art} published a book entitled ''The art and science of analyzing software data'', which is driven by the MSR (Mining Software Repositories) community and focuses mainly on data analysis based on statistics and machine learning. Another related book published by \cite{menzies2016perspectives} covers perspectives on data science for software engineering by various authors.

\cite{kitchenham2015evidence} published a book entitled ''Evidence-based software engineering and systematic reviews'', which provides practical guidance on how to conduct secondary studies in software engineering. The book also discusses the nature of evidence and explains the types of primary studies that provide inputs to a secondary study.

\cite{malhotra2016empirical} published a book entitled ''Empirical research in software engineering: concepts, analysis, and applications'', which shows how to implement empirical research processes, procedures, and practices in software engineering. The book covers many accompanying exercises and examples. The author especially also discusses the process of developing predictive models, such as defect prediction and change prediction, on data collected from source code repositories, and, more generally the application of machine learning techniques in empirical software engineering. 

\cite{ben2017empirical} published a book entitled ''Empirical Research for Software Security: Foundations and Experience'', which discusses empirical methods with a special focus on software security.

\cite{staron2019empirical} published a book entitled "Action Research in Software Engineering: Theory and Applications" , which offers a comprehensive discussion on the use of action research as an instrument to evolve software technologies and promote synergy between researchers and practitioners. 

In addition to these textbooks, there are also edited books available that are related to special events in empirical software engineering and cover valuable methodological contributions. 

\cite{rombach1993experimental} edited proceedings from a Dagstuhl seminar in 1992 on empirical software engineering entitled ''Experimental Software Engineering Issues: Critical Assessment and Future Directions''. The goal was to discuss the state of the art of empirical software engineering by assessing past accomplishments, raising open questions, and proposing a future research agenda at that time. However, many contributions of that book are still relevant today.

\cite{conradi2003empirical} edited a book entitled ''Empirical methods and studies in software engineering: experiences from ESERNET'', which covers experiences from the Experimental Software Engineering Research NETwork (ESERNET), a thematic network funded by the European Union between 2001 and 2003.

\cite{boehm2005foundations} edited a book entitled ''Foundations of empirical software engineering: the legacy of Victor R. Basili'' on the occasion of V. R. Basili's 65th birthday, which covers reprints of 20 papers that defined much of his work.

\cite{basili2007empirical} edited proceedings from another Dagstuhl seminar in 2006 on empirical software engineering entitled ''Empirical Software Engineering Issues. Critical Assessment and Future Directions''. 

\cite{munch2013perspectives} edited a book entitled ''Perspectives on the future of software engineering: essays in honor of Dieter Rombach'' on the occasion of Dieter Rombach's 60th birthday, which covers contributions by renowned researchers and colleagues of him.

\section{Conclusion}
\label{sec:conclusion}

In this chapter we presented the evolution of empirical software engineering in five iterations, i.e., (1) mid-1960s to mid-1970s, (2) mid-1970s to mid-1980s, (3) mid-1980s to end of the 1990s, (4) the 2000s, and (5) the 2010s. We presented the five iterations of the development of empirical software engineering mainly from a methodological perspective and additionally took key papers, venues, and books into account. Available books explicitly dedicated to empirical research methodology in software engineering were covered in chronological order in a separate section on recommended further readings. Furthermore, we discuss -- based on the chapters in this book -- trends on contemporary empirical methods in software engineering related to the plurality of research methods, human factors, data collection and processing, aggregation and synthesis of evidence, and impact of software engineering research. 

\begin{acknowledgement}
We thank all the authors and reviewers of this book on contemporary empirical methods in software engineering for their valuable contribution 
\end{acknowledgement}
\bibliographystyle{spbasic}
\bibliography{references}

\end{document}